# Narrow band perfect absorber for maximum localized magnetic and electric field enhancement and sensing applications

Zhengdong Yong[1], Senlin Zhang[1], Chengsheng Gong[1], and Sailing He[*, 1, 2]

[1]Centre for Optical and Electromagnetic Research, State Key Laboratory of Modern Optical Instrumentations, Zhejiang University, Hangzhou 310058, China.

[2]Department of Electromagnetic Engineering, School of Electrical Engineering, Royal Institute of Technology (KTH), S-100 44 Stockholm, Sweden

* Email for the corresponding author: sailing@kth.se

**Plasmonics offer an exciting way to mediate the interaction between light and matter, allowing strong field enhancement and confinement, large absorption and scattering at resonance. However, simultaneous realization of ultra-narrow band perfect absorption and electromagnetic field enhancement is challenging due to the intrinsic high optical losses and radiative damping in metals. Here, we propose an all-metal plasmonic absorber with an absorption bandwidth less than 8nm and polarization insensitive absorptivity exceeding 99%. Unlike traditional Metal-Dielectric-Metal configurations, we demonstrate that the narrowband perfect absorption and field enhancement are ascribed to the vertical gap plasmonic mode in the deep subwavelength scale, which has a high quality factor of 120 and mode volume of about $10^{-4} \cdot (\lambda_{res}/n)^3$. Based on the coupled mode theory, we verify that the diluted field enhancement is proportional to the absorption, and thus perfect absorption is critical to maximum field enhancement. In addition, the proposed perfect absorber can be operated as a refractive index sensor with a sensitivity of 885nm/RIU and figure of merit as high as 110. It provides a new design strategy for narrow band perfect absorption and local field enhancement, and has potential applications in biosensors, filters and nonlinear optics.**

Resonant plasmonic and metamaterial nanostructures have attracted much attention in the past decade due to their exotic dynamic properties that are not available in nature, such as optical negative refraction[1,2], perfect lensing[3,4] and electromagnetic cloaking[5]. Collective oscillations of free electrons in metals, known as localized or delocalized surface plasmons[6] lie at the origin of these unique properties, allowing a multitude of exciting applications such as biosensors[7-9], optical filters[10], photodetectors[11] and nanolasers[12]. While the intrinsic optical loss of metals is a major limitation in the performance of these devices, it is advantageous for enhancing light absorption. In 2008, Landy *et al.*[13] first proposed a perfect metamaterial absorber with nearly perfect absorbance by simultaneously exciting electric and magnetic resonances to realize the impedance match with the surrounding air. After that, substantial absorbers based on different physical mechanisms have been demonstrated theoretically and experimentally in a wide spectral range, which can be generally categorized into broadband absorbers[14-16] and narrowband absorbers[17-21] in terms of their absorption bandwidth. While broadband absorbers are generally used in thermo-photovoltaics[22], narrowband perfect absorbers can be used in sensing[9, 19, 20], absorption filtering[23] and thermal radiation tailoring[24, 25].

For sensing applications, Liu *et al.*[9] experimentally realized a refractive index sensor by using an infrared perfect absorber, indicating that both narrow bandwidth and large modulation depth are necessary to improve the sensing performance. The triple-layer metal-dielectric-metal (MDM) configuration they used has been widely applied to the previous plasmonic absorbers, where a thin dielectric spacer is used to enable strong plasmonic coupling between the top resonators and the bottom metal film. Such an absorber design can also be intuitively treated as a resonator coupled to a single input transmission line, with the dielectric spacer thickness influencing the radiative damping rates and resonant frequency[26]. However, due to the strongly radiative damping and the inherent metal loss, the resonant absorption bandwidths of these plasmonic absorbers are relatively broad (>40nm), which severely hampers its applications. Thus, it is of great significance to design ultra-narrow band perfect plasmonic absorbers. Up to date, several theoretical and experimental efforts have been devoted to achieve this. Among them, Li *et al.*[18] experimentally realized a narrow band absorber with an absorption bandwidth of 12nm and absorptivity exceeding 90% based on surface lattice resonance. Ultra-narrow band perfect absorbers based on a plasmonic analog of electromagnetically induced absorption and optimized grating were theoretically proposed[19, 20]. In [21], the authors designed a nanoslit-microcavity-based narrow band absorber with bandwidth of 8nm and sensing figure of merit (FOM) of 25, which is much higher than the previous sensor with FOM less than 10[17, 27, 28].

In a different context, metal nanostructures based on localized surface plasmonic resonance have generally been effective in creating strongly enhanced electromagnetic fields. Termed electric hot spots, the confined electric field in metal particle junctions enables large enhancement of emission processes and nonlinearities[6], which are mainly mediated by the electric polarization of molecules. In particular, magnetic activity at optical frequencies is far smaller than its electric counterpart because of the extremely weak magnetic response of natural materials. As a result, magnetic hot spots are highly desirable for strengthening the magnetic response. Many structures have been specifically designed to achieve magnetic hot spots, such as diabolo antennas[29], two parallel metal plates[30] and closely spaced thick gold rings[31]. However, achieving simultaneous electric and magnetic hot spots at the same spatial position is rather challenging. Moreover, to maximize the field enhancement, coupled mode theory has recently been used[32, 33]. Both the quality factor and mode volume are critical parameters in engineering the local field enhancement, while it seems separate between the far-field perfect absorption and the maximum near-field enhancement.

In this paper, we propose an all-metal absorber with an absorption bandwidth less than 8nm and polarization insensitive absorptivity exceeding 99%. Full-wave electromagnetic simulations reveal that the narrowband perfect absorber with quality factor of 120 can simultaneously create giant electric and magnetic field enhancements in the deep subwavelength scale (mode volumes $\cong 10^{-4} \cdot (\lambda/n)^3$). We further demonstrate that perfect absorption is necessary to maximize the local field enhancement based on the coupled mode theory. Additionally, operated as a refractive index sensor, the proposed absorber has a high sensitivity of 885nm/RIU and figure of merit (FOM) up to 110 in the near infrared region. Our findings provide a new method to engineer narrowband perfect absorption and local field enhancement. It is expected that such absorber structures will hold great potential in sensing and near field optics.

## Results

**Structure and parameters.** The schematic of the proposed structure is depicted in Fig. 1 with a magnified unit cell (enclosed in dashed box) shown in the inset. The all-metal structure consists of periodic arrays of coupled thick silver disks placed directly on the surface of a uniform silver film. The closely spaced silver disks have a radius ($r$) of 80*nm*, thickness ($t$) of 100*nm* and gap distance ($g$) of 20*nm*. The period constant ($p$) of the arrays is 470*nm* and the bottom silver film has a thickness of 100*nm*. In addition, the whole structure is placed on a glass substrate, and the surrounding material is assumed to be air. The complex dielectric constant of silver is modeled by a Drude-Lorentz fitting (with 6 coefficients) to tabulate experimental data[34], and the permittivity of the glass is 2.25. These structures are compatible with the current fabrication technology such as electron beam lithography and focused ion milling.

**Ultra-narrow band perfect absorption based on vertical gap plasmonic mode.** Fig. 2(*a*) depicts the absorption and reflection spectrum of the all-metal structures. The spectral absorption (*A*) is defined by *A=1-R-T*, where *R* and *T* are the reflection and transmission of the structures, respectively. Since the thickness of the bottom silver film is thicker than the skin depth in the infrared region, the transmission channel is prevented and the absorption is reduced to *A=1-R*. As shown in Fig. 2(*a*), there is a distinctive resonance at the wavelength of 918*nm*, with a bandwidth (or *FWHW*) of about 7.5*nm* and absorptivity of 99.6%, which is far narrower than those of MDM based perfect absorbers, whose bandwidths are larger than 40*nm*. Due to the symmetry of the structures, polarization-insensitive absorption can be easily achieved.

To reveal the physical mechanism in the proposed perfect absorber, we calculate the electric, magnetic field ($|E_x|$ and $|H_y|$) and current density distributions at the resonant wavelength of 918nm, and map the absorbed power density in both the x-y and x-z planes of the structure in Fig. 2(*b*). It is evident that both the electric and magnetic field are strongly concentrated in the gap region as well as the absorbed power density. The resonance oscillates like a magnetic dipole, which can be seen from the confined magnetic field in the gap and the arrow line distribution of the current density in the x-z plane. A vertical Fabry-Perot cavity of MDM waveguide is formed by the disks, and the air gap acts as the dielectric layer[35, 36]. With the help of the silver film, this resonance can be well excited. Therefore, we can attribute the narrow band perfect absorption to the vertical gap plasmonic mode, due to the relatively poor scattering ability of the magnetic dipole resonance in the deep subwavelength region. To better understand the properties of the vertical gap plasmonic mode, the influences of the radius and thickness of the disks on the absorption spectrum are investigated. The corresponding results are shown in Fig. 2(c) and (d). As the radius of the disks decreases from 100nm to 60nm, the resonant wavelength is slightly shifted, which is consistent with previous studies[35]. By contrast, the resonant wavelength is sensitive to the thickness of the disks, which redshifts from 809nm to 1087nm as the thickness is increased from 80nm to 130nm. This can be intuitively explained by an increase in the vertical cavity length.

**Maximum electromagnetic field enhancement based on coupled mode theory.** Due to the localized surface plasmonic resonance, plasmonic nanostructures can efficiently link far field radiation with the localized near field, and can be viewed as optical nanoantennas[37]. They are specifically designed to produce large electromagnetic field enhancement within a small mode volume, called a hot spot. In particular, magnetic hot spots are highly desirable for enhancing magnetic dipole transitions and developing magnetic-based devices due to the extremely weak magnetic response of the natural material in the optical domain[31], while electric hot spots play a crucial role in SERS, nanolasers and nonlinear optics. Before numerically examining the field enhancement, we first refer to the temporal coupled mode theory which has been used previously to model the field enhancement of an optical antenna[32, 33]. Consider the structures as optical antennas with an effective radiation cross section of $A_c$, illuminated by an incident beam of cross section $A_i$. The maximum field enhancement at resonance can be expressed[33] (Supplementary Section S1 online) as $\left.\frac{|E_{loc}|^2}{|E_i|^2}\right|_{max} = \frac{A_i \lambda_{res}}{\pi} \frac{Q}{V_{eff}}$, where $E_{loc}$ and $E_i$ are the local electric field in the gap region and the incident electric field, respectively. $Q$ is the quality factor, and $V_{eff}$ is the effective mode volume of the resonator. The maximum field enhancement is proportional to the Purcell factor[38] ($Q/V_{eff}$), which can be achieved when simultaneously the radiative decay rate $\gamma_r$ equals the absorption decay $\gamma_a$ and the antenna's radiation pattern matches the incident beam shape, denoted by $\gamma_r = \gamma_a$, and $A_c = A_i$. In addition, the critical coupling condition leads to perfect absorption[26] (see Supplementary Section S2 online).

The electric and magnetic field intensity enhancement measured at a point in the middle of the gap and at a height $t/2$ over the silver film are shown in Fig. 3(*a*), and the electric field intensity enhancement at the top surface (height $t$) is also plotted. Both the narrowband electric and magnetic fields are strongly enhanced in the same spatial region (both E and B enhancements are relatively large at a height $t/2$ over the silver film, and their hot spots have spatial overlap at around t/2). This can be ascribed to the vertical gap plasmonic mode. From Fig. 2(b), one sees that the currents (the arrow line distribution in the x-z plane) flow vertically between the thick silver disks through the bottom connected silver film, creating a current loop which gives rise to one-order higher magnetic field than that which would result from putting on the dielectric substrate[31]. This is much like the vertical SRRs[39]. Meanwhile, the narrow bandwidth confirms the mode's high quality factor $Q$ to be as high as 120, much higher than those of 20 in the MDM based structures[9, 33] and 79 in recently reported ultra-narrow absorbers based on surface lattice resonances[18]. It simultaneously supports a mode volume $V$ of $10^{-4} \cdot (\lambda_{res}/n)^3$ at the resonant wavelength of 918nm in the surrounding material ($n$=1), which is much smaller than the traditional MDM structures. The high quality factor $Q$ and extremely small $V$ lead to the observed large electric field enhancement based on the above coupled mode theory.

To validate the relation between field intensity enhancement and absorption, we only change the period constant $p$ and keep other parameters the same. This will intuitively influence the ratio of the hot-spot area to the unit cell size, and thus we dilute the field intensity enhancement with the ratio of the hot-spot area to unit cell size as $\frac{|E_{norm}|^2}{|E_i|^2} = \frac{S_{silver}}{p^2} \cdot \frac{|E_{loc}|^2}{|E_i|^2}$, where $S_{silver}$ is the total top-surface area of the four silver disks in a unit cell. Such a diluted field intensity enhancement is useful to evaluate the performance of multi-hot spot devices[40, 41]. Fig. 3(b) shows the diluted field intensity enhancement and absorption plots as the period *p* varies. We find that the diluted field intensity enhancement is approximately proportional to the absorption, which is consistent with the coupled mode theory, except when the period constant is so small that the near field coupling becomes noticeable and much energy is dissipated between unit cells. Furthermore, the resonant wavelength is slightly shifted as seen from the inset of Fig. 3(b). Therefore, to maximize the local field enhancement, the absorption is more easily engineered compared with the *Q*-matching condition[33], and perfect absorption is critical.

**Plasmon sensing capability of the structure.** As is well known, the resonant wavelength of plasmonic nanostructures is dependent on the refractive index of the surrounding dielectric environment[27], a property that has been widely utilized for sensing applications. The sensing capability is usually described by the following definitions of sensitivity and figure of merit (FOM)[17, 21].

$$S = \frac{\Delta \lambda}{\Delta n}, FOM = \frac{S}{FWHW}; S^* = \frac{\Delta I}{\Delta n}, FOM^* = \frac{S^*}{I}$$

where $\Delta\lambda$ is the spectral shift caused by a certain refractive index change in the environment $\Delta n$, $\Delta I$ is the detected intensity change for a particular incident wavelength, and *I* is the absolute intensity. Since our all-metal structure has a narrow bandwidth and nearly zero reflectance (R=0.44%) around the resonant frequency, it is expected to have good sensing capability. To demonstrate its performance, we vary the surrounding refractive index from 1 to 1.05 with a step interval of 0.01, and the corresponding reflection spectra are displayed in Fig. 4(a), where obvious redshift of the resonance is observed. The slope of the curve in Fig. 4(b) represents the sensitivity *S* of 885nm/RIU, and an *FOM* of 110 can be achieved considering its narrow bandwidth. This is much higher than other recently reported values[21, 27, 28, 39]. Furthermore, as seen from Fig. 4(a), a slight spectral shift will cause a large optical intensity variation. We can obtain $S^*$=85/RIU and $FOM^*$=19000 at a fixed measurement wavelength of 918nm from the definition because of the near-unity absorption, and the $S^*$ is about one order larger than that of the cavity enhanced localized plasmonic resonance sensing[17].

**Discussions**

In summary, we have proposed an all-metal absorber with an absorption bandwidth less than 8nm and polarization insensitive absorptivity exceeding 99%. The absorber is based on the localized vertical gap plasmonic mode, unlike traditional MDM configurations. Due to the high quality factor and extremely small mode volume, the perfect absorber can achieve both electric and magnetic hot spots at the same position based on coupled mode theory. We have investigated the relation between the far-field absorption and the localized near-field enhancement, and found that the perfect absorption is critical to maximizing the field enhancement besides the previous $Q$-matching condition. Considering the superior localized plasmonic characteristic of the mode, we demonstrated its sensing capability, and a high sensitivity of 885nm with FOM up to 110 has been realized. This is much better than most reported values. Our structures can be well tuned over the infrared domain by changing the structure parameters. These findings provide a new design strategy not only for narrow band perfect absorbers but also near field engineering, both electric and magnetic. Such narrowband resonators will easily find applications in thermo-photovoltaics, biosensors, nonlinear plasmonics and lasers.

**Methods**

**Simulation.** Three-dimensional finite-difference time-domain calculations were performed using a commercially available software package (Lumerical FDTD Solutions Inc.v8.6)[42]. Due to the symmetry of the structures, only the plane wave polarized along the x-axis is considered as the excitation source and is incident from the top. Periodic boundary conditions are employed for the lateral boundaries, and perfectly matching layers are applied along the z direction to eliminate the boundary scattering. The mesh size is set to be 0.5 nm which is much smaller than the element sizes and the operating wavelength, and a standard convergence test is done to ensure negligible numerical errors. 2D frequency-domain field and power monitors (perpendicular to the x-y plane) are used to calculate the reflection and transmission, and point monitors are used to record the electromagnetic field enhancement.

**Quality factor and Mode volume.** The quality factor was estimated by $Q = \lambda_{res}/\Delta\lambda$, where $\lambda_{res}$ is the resonance wavelength, and $\Delta\lambda$ is spectral width. The mode volume of the proposed structures was calculated using the formula $V = \int \frac{\varepsilon(r)|E(r)|^2}{\max(\varepsilon(r)|E(r)|^2)} d^3r$, where $\varepsilon(r)$ is the complex dielectric constant at position $r$, and $|E(r)|^2$ is the corresponding electric field intensity[32]. We first calculate the field distribution in the whole structure by using FDTD and then the mode volume $V$.

**References**

1. Shelby, R. A., Smith, D. R., & Schultz, S. Experimental verification of a negative index of refraction. *Science* **292**, 77-79 (2001).

2. Shalaev, V. M. Optical negative-index metamaterials. *Nature Photon*. **1**, 41-48 (2007).

3. Pendry, J. B. Negative refraction makes a perfect lens. *Phys. Rev. Lett*. **85**, 3966 (2000).

4. Fang, N., Lee, H., Sun, C. & Zhang, X. Sub-diffraction-limited optical imaging with a silver superlens. *Science* **308**, 534 (2005).

5. Schurig, D., Mock, J. J., Justice, B. J., Cummer, S. A., Pendry, J. B., Starr, A. F., & Smith, D. R. Metamaterial electromagnetic cloak at microwave frequencies. *Science* **314**, 977-980 (2006).

6. Maier, S. A. *Plasmonics: fundamentals and applications* (Springer, 2007).

7. Anker, J. N., Hall, W. P., Lyandres, O., Shah, N. C., Zhao, J., & Van Duyne, R. P. Biosensing with plasmonic nanosensors. *Nature Mater*. **7**, 442-453 (2008).

8. Kabashin, A. V., Evans, P., Pastkovsky, S., Hendren, W., Wurtz, G. A., Atkinson, R., & Zayats, A. V. Plasmonic nanorod metamaterials for biosensing. *Nature Mater*. **8**, 867-871 (2009).

9. Liu, N., Mesch, M., Weiss, T., Hentschel, M., & Giessen, H. Infrared perfect absorber and its application as plasmonic sensor. *Nano Lett*. **10**, 2342-2348 (2010).

10. Ellenbogen, T., Seo, K., & Crozier, K. B. Chromatic plasmonic polarizers for active visible color filtering and polarimetry. *Nano Lett*. **12**, 1026-1031 (2012).

11. Sobhani, A., Knight, M. W., Wang, Y., Zheng, B., King, N. S., Brown, L. V., & Halas, N. J. Narrowband photodetection in the near-infrared with a plasmon-induced hot electron device. *Nature Comm.* **4**, 1643 (2013).

12. Oulton, R. F., Sorger, V. J., Zentgraf, T., Ma, R. M., Gladden, C., Dai, L., & Zhang, X. Plasmon lasers at deep subwavelength scale. *Nature* **461**, 629-632 (2009).

13. Landy, N. I., Sajuyigbe, S., Mock, J. J., Smith, D. R., & Padilla, W. J. Perfect metamaterial absorber. *Phys. Rev. Lett*. **100**, 207402 (2008).

14. Aydin, K., Ferry, V. E., Briggs, R. M., & Atwater, H. A. Broadband polarization-independent resonant light absorption using ultrathin plasmonic super absorbers. *Nature Commun*. **2**, 517 (2011).

15. Ding, F., Cui, Y., Ge, X., Jin, Y., & He, S. Ultra-broadband microwave metamaterial absorber. *Appl. Phys. Lett*. **100**, 103506 (2012).

16. Cui, Y., Fung, K. H., Xu, J., Ma, H., Jin, Y., He, S., & Fang, N. X. Ultrabroadband light absorption by a sawtooth anisotropic metamaterial slab. *Nano Lett*. **12**, 1443-1447 (2012).

17. Ameling, R., Langguth, L., Hentschel, M., Mesch, M., Braun, P. V., & Giessen, H. Cavity-enhanced localized plasmon resonance sensing. *Appl. Phys. Lett*. **97**, 253116 (2010).

18. Li, Z., Butun, S., & Aydin, K. Ultranarrow band absorbers based on surface lattice resonances in nanostructured metal surfaces. *ACS Nano* **8**, 8242-8248 (2014).

19. Meng, L., Zhao, D., Ruan, Z., Li, Q., Yang, Y., & Qiu, M. Optimized grating as an ultra-narrow band absorber or plasmonic sensor. *Opt. Lett*. **39**, 1137-1140 (2014).

20. He, J., Ding, P., Wang, J., Fan, C., & Liang, E. Ultra-narrow band perfect absorbers based on plasmonic analog of electromagnetically induced absorption. *Opt. Express* **23**, 6083-6091 (2015).

21. Lu, X., Zhang, L., & Zhang, T. Nanoslit-microcavity-based narrow band absorber for sensing applications. *Opt. Express* **23**, 20715-20720 (2015).

22. Atwater, H. A., & Polman, A. Plasmonics for improved photovoltaic devices. *Nature Mater*. **9**, 205-213 (2010).

23. Lee, K. T., Seo, S., & Guo, L. J. High-color-purity subtractive color filters with a wide viewing angle based on plasmonic perfect absorbers. *Adv. Opt. Mater*. **3**, 347-352 (2015).

24. Liu, X., Tyler, T., Starr, T., Starr, A. F., Jokerst, N. M., & Padilla, W. J. Taming the blackbody with infrared metamaterials as selective thermal emitters. *Phys. Rev. Lett*. **107**, 045901 (2011).

25. Song, M., Yu, H., Hu, C., Pu, M., Zhang, Z., Luo, J., & Luo, X. Conversion of broadband energy to narrowband emission through double-sided metamaterials. *Opt. Express* **21**, 32207-32216 (2013).

26. Wu, C., Burton Neuner, I. I. I., Shvets, G., John, J., Milder, A., Zollars, B., & Savoy, S. Large-area wide-angle spectrally selective plasmonic absorber. *Phys. Rev. B* **84**, 075102 (2011).

27. Becker, J., Trügler, A., Jakab, A., Hohenester, U., & Sönnichsen, C. The optimal aspect ratio of gold nanorods for plasmonic bio-sensing. *Plasmonics* **5**, 161-167 (2010).

28. Huang, C., Ye, J., Wang, S., Stakenborg, T., & Lagae, L. Gold nanoring as a sensitive plasmonic biosensor for on-chip DNA detection. *Appl. Phys. Lett*. **100**, 173114 (2012).

29. Grosjean, T., Mivelle, M., Baida, F. I., Burr, G. W., & Fischer, U. C. Diabolo nanoantenna for enhancing and confining the magnetic optical field. *Nano Lett*. **11**, 1009-1013 (2011).

30. Cai, W., Chettiar, U. K., Yuan, H. K., de Silva, V. C., Kildishev, A. V., Drachev, V. P., & Shalaev, V. M. Metamagnetics with rainbow colors. *Opt. Express* **15**, 3333-3341 (2007).

31. Lorente-Crespo, M., Wang, L., Ortuño, R., García-Meca, C., Ekinci, Y., & Martínez, A. Magnetic hot spots in closely spaced thick gold nanorings. *Nano Lett*. **13**, 2654-2661 (2013).

32. Maier, S. A. Plasmonic field enhancement and SERS in the effective mode volume picture. *Opt. Express* **14**, 1957-1964 (2006).

33. Seok, T. J., Jamshidi, A., Kim, M., Dhuey, S., Lakhani, A., Choo, H., & Yablonovitch, E. Radiation engineering of optical antennas for maximum field enhancement. *Nano Lett*. **11**, 2606-2610 (2011).

34. Johnson, P. B., & Christy, R. W. Optical constants of the noble metals. *Phys. Rev. B*, **6**, 4370 (1972).

35. Bozhevolnyi, S. I., & Søndergaard, T. General properties of slow-plasmon resonant nanostructures: nano-antennas and resonators. *Opt. Express* **15**, 10869-10877 (2007).

36. Le Perchec, J., Quemerais, P., Barbara, A., & Lopez-Rios, T. Why metallic surfaces with grooves a few nanometers deep and wide may strongly absorb visible light. *Phys. Rev. Lett*. **100**, 066408 (2008).

37. Novotny, L., & Van Hulst, N. Antennas for light. *Nature Photon*. **5**, 83-90 (2011).

38. Purcell, E. M. Spontaneous transition probabilities in radio-frequency spectroscopy. *Phys. Rev*. **69**, 681 (1946).

39. Wu, P. C., Sun, G., Chen, W. T., Yang, K. Y., Huang, Y. W., Chen, Y. H., & Tsai, D. P. Vertical split-ring resonator based nanoplasmonic sensor. *Appl. Phys. Lett*. **105**, 033105 (2014).

40. Chu, Y., Banaee, M. G., & Crozier, K. B. Double-resonance plasmon substrates for surface-enhanced Raman scattering with enhancement at excitation and stokes frequencies. *ACS Nano* **4**, 2804-2810 (2010).

41. Linden, S., Niesler, F. B. P., Förstner, J., Grynko, Y., Meier, T., & Wegener, M. Collective effects in second-harmonic generation from split-ring-resonator arrays. *Phys. Rev. Lett*. **109**, 015502 (2012).

42. Lumerical Solutions, Inc. FDTD SOLUTIONS. Available at: http://www.lumerical.com.

**Acknowledgement**

This work was partially supported by the National Natural Science Foundation of China (Nos. 91233208, 61271016 and 61178062), the National High Technology Research and Development Program (863 Program) of China (No. 2012AA030402), the Program of Zhejiang Leading Team of Science and Technology Innovation, the Postdoctoral Science Foundation of China (No. 2013M541774), the Preferred Postdoctoral Research Project Funded by Zhejiang Province (No. BSH1301016), and Swedish VR grant (No. 621-2011-4620).

**Author contributions**

Z.D. Y. performed the simulation and analyzed the data. S.L. Z. and C.S. G. gave valuable discussions and contributed to the manuscript preparation. S. H. supervised the whole work. The writing of the manuscript was done by Z.D. Y. and S. H., and finalized by S. H.

**Additional Information**

**Supplementary information** accompanies this paper at http://www.nature.com/scientificreports
**Competing financial interests:** The authors declare no competing financial interests.

**Figures and legends**

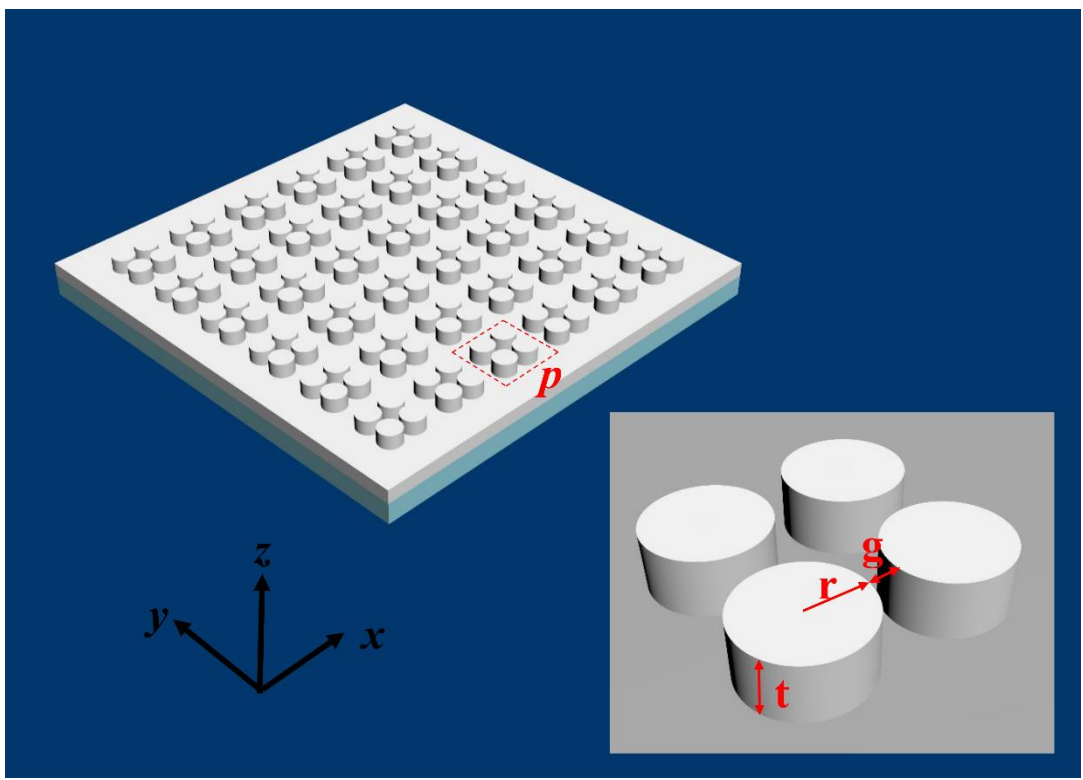

Fig. 1 A schematic diagram of the all-metal perfect absorber with a magnified unit cell (enclosed in dashed box) shown in the inset: periodic arrays of coupled thick silver disks are placed directly on the surface of a uniform silver film. The closely spaced silver disks have a radius ($r$) of 80$nm$, thickness ($t$) of 100$nm$ and gap distance ($g$) of 20$nm$. The period constant ($p$) of the arrays is 470$nm$ and the bottom silver film has a thickness of 100$nm$. In addition, the whole structure is placed on a glass substrate, and the surrounding material is assumed to be air.

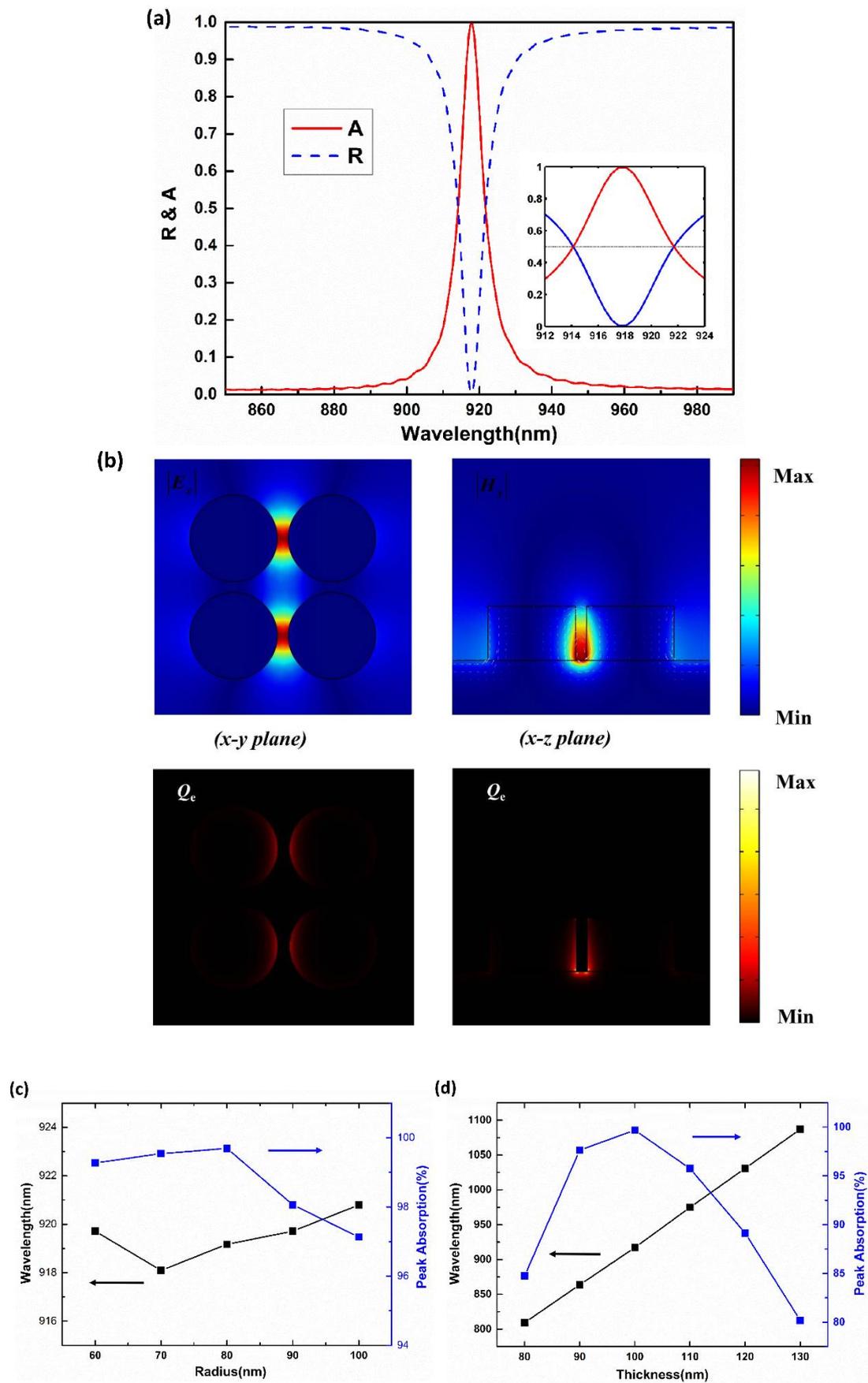

Fig. 2 (a) The absorption and reflection spectrum with magnified spectrum shown in the inset. (b) Distributions of the electric field $|E_x|$ (color bar in the x-y plane), magnetic field $|H_y|$ (color bar in the x-z plane) and current density(arrow line in the x-z plane) at resonance (top row), and mapping of the absorbed power density (bottom row) in both the x-y and x-z planes. Peak absorption and resonant wavelength as the radius (c) and thickness (d) of the structures varies.

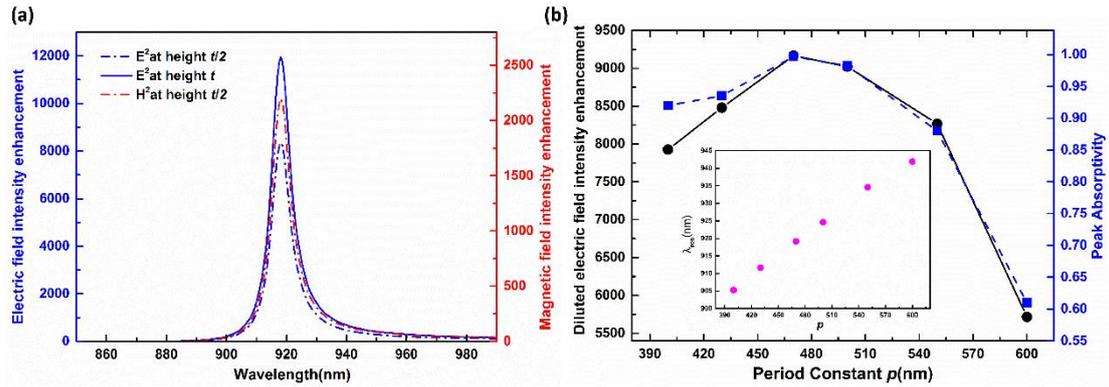

Fig. 3 (a) Electric (blue dash-dot curve) and magnetic field (red dash-dot curve) intensity enhancement at a point in the middle of the gap and height t/2 over the silver film, and electric field intensity enhancement at height t (blue curve). (b) Diluted electric field intensity enhancement (circles connected with solid lines) and the peak absorptivity (squares connected with dashed lines) as the period varies. Inset: resonant wavelength vs period with other parameters unchanged.

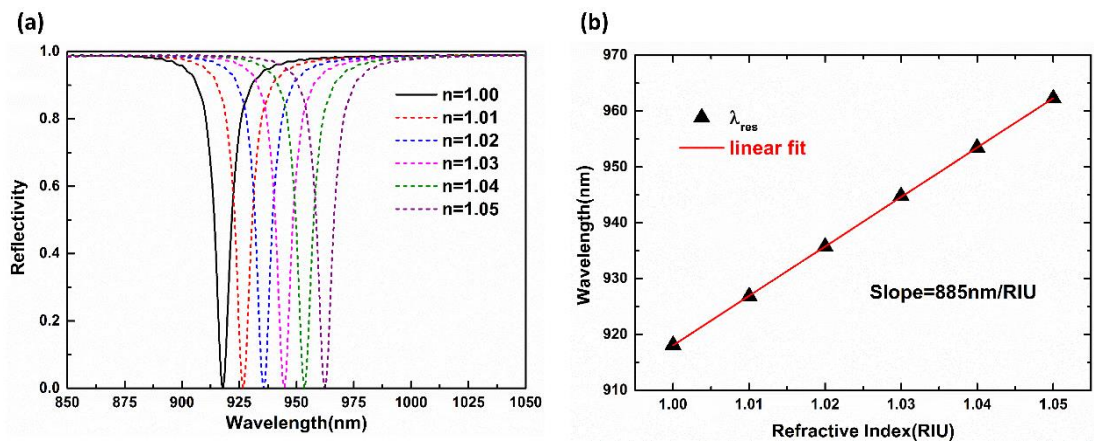

Fig. 4 (a) Reflection spectrum of the all-metal perfect absorber with the refractive index varying from 1 to 1.05 with a step interval of 0.01. (b) Resonant wavelength as a function of the surrounding low refractive index. The red line is the linear fitting with the slope representing the sensitivity $S$.